# The Future of Spreadsheets in the Big Data Era

David Birch[1*], David Lyford-Smith[2] & Yike Guo[1]
[1]Data Science Institute, Imperial College London
[2]ICAEW IT Faculty
[*]david.birch@imperial.ac.uk

**ABSTRACT**
*The humble spreadsheet is the most widely used data storage, manipulation and modelling tool. Its ubiquity over the past 30 years has seen its successful application in every area of life. Surprisingly the spreadsheet has remained fundamentally unchanged over the past three decades. As spreadsheet technology enters its 4$^{th}$ decade a number of drivers of change are beginning to impact upon the spreadsheet. The rise of Big Data, increased end-user computing and mobile computing will undoubtedly increasingly shape the evolution and use of spreadsheet technology.*

*To explore the future of spreadsheet technology a workshop was convened with the aim of "bringing together academia and industry to examine the future direction of spreadsheet technology and the consequences for users". This paper records the views of the participants on the reasons for the success of the spreadsheet, the trends driving change and the likely directions of change for the spreadsheet. We then set out key directions for further research in the evolution and use of spreadsheets. Finally we look at the implications of these trends for the end users who after all are the reason for the remarkable success of the spreadsheet.*

## 1. INTRODUCTION

The goal of this paper is to explore the future of the spreadsheet in the light of the age of Big Data. Firstly we explore the reasons underlying the success of spreadsheet technologies over the past 3 decades before exploring the challenges encountered using current spreadsheet technologies, many of these such as high error rates are well-known, at least within the research community.

As computing continues to evolve we then explore the "Drivers of Change" which the authors believe are likely to impact upon the development and use of the future spreadsheet. Key amongst these drivers is the rise of Big Data and Artificial Intelligence (AI) which will be explored in depth. These drivers are likely to result in particular "Directions of Change" specific to the evolution of spreadsheets. Key among these directions of change include better support for dealing with large external datasets and the ability to integrate such data together.

How spreadsheet technology should adapt in response to drivers of change presents a range of open research questions, we outline the most compelling research questions in section 6. Finally we conclude this paper with a look at the implications for spreadsheet users who after all have made the spreadsheet such a success and will drive spreadsheets success far into the future.

## 2. THE SUCCESS OF THE SPREADSHEET

Before commenting on the future of the spreadsheet it is necessary to explore the factors underlying the spreadsheets enduring success. From literature, discussions during the workshop and the author's experiences, we identify the following factors:

### 2.1 Ubiquity
Spreadsheets are universally available to students and professionals, indeed most computer literate people have a familiarity with spreadsheets. Spreadsheets exist on most operating systems, architectures and form factors, this has made them a common tool and format for working with and exchanging data, encoding calculations and modelling business situations.

### 2.2 Unconstrained canvas

A blank spreadsheet is just that: blank. It may become a shopping list, a financial model or a 250,000 cell multi-disciplinary model. This flexibility, while it is the spreadsheet's biggest asset, frequently, proves it's Achilles' heel, as we will highlight. This adaptability has led to spreadsheets being applied in nearly every industry and situation. One of the key reasons for this is the openness of the spreadsheet and the clarity of its mental model – all a user needs to understand is the concept of tables of cells which may contain text, numbers or dates or may be formulas calculating with such data.

### 2.3 An open box
Unlike many business applications, spreadsheets are inherently open, there is no "black box"; everything is open and can, *in theory*, be understood and changed cell by cell. This, of course, does not always happen in practice as user's self-confidence and the sheer complexity of large formulas and models may overwhelm and are known to cause hidden errors in spreadsheets [Panko 2015].

### 2.4 Ability to communicate data
Spreadsheets provide simple-to-use data visualisation tools which enable communication of quantitative insight succinctly to a wide audience. They also provide an early way of understanding and exploring data to discover trends and patterns. Since these visualisations are often based on the results of calculated data, they turn into an interactive graphical tool for visually answering questions about a modelled case study. As spreadsheets are a standard format these interactive models or summaries of data can be shared visually to a wide audience.

### 2.5 Ability to store data in a computable format
For many end users the primary use of a spreadsheet is to store data. A spreadsheet's expandable tabular format enables extensible storage of data. Key to their utility is this computability of this format, from a simple sum of a table to a depreciation calculation to a mail merge, making data computable is a key feature behind the success of spreadsheets.

### 2.6 Lightweight databases
Coupled with the ability to store data in a computable format we note that the spreadsheets often act as a lightweight database. This is particularly apparent for users without training in or access to more advanced database systems. Recently analysis of a large companies spreadsheets showed that one third of files contained only data and no formulas [Hermans 2015]. Spreadsheets allow the storage of data in a format which allows the look up of data items, the filtering of data to meet criteria and the "joining" of data tables which are the basic features of a database. This feature trend is increasing with systems such as Microsoft's Power Pivot providing tools to summarise data tables graphically data directly as if the user were writing in a query language such as SQL.

### 2.7 Ability to model and test scenarios
Another common use of spreadsheets is to act as a computable "back of the envelope" for carrying out calculations to model scenarios. Once created these models are then able to be used to test a variety of scenarios with the same calculation logic. This enables users to gain an understanding the possible design space of outcomes. The spreadsheet's ability to encode complex logical, statistical or engineering calculations assists users to carry out tasks which would be challenging manually or would otherwise require customised software or the ability to program.

### 2.8 Accessible End-User Programming
Spreadsheets are widely regarded as the most common means of end user programming [Scaffidi 2005]. The simple mental model and "what you see is what you get" environment has enabled many millions of people to create complex calculation algorithms implemented naturally and incrementally. This has enabled spreadsheets to be adapted to many specific scenarios.

### 2.9 Advanced Functionality
A key feature enabling the spreadsheet's success is its ability to grow with the end-users competence. Initially the model of spreadsheet appears simple it may be augmented by numerous complex formulas and algorithms. From experimenting with logical and lookup functions through the creation of Macros to the writing of complex scripts in language such as Visual Basic for Applications, most spreadsheet systems provide advanced functionality to enable users to accomplish complex tasks.

**2.10 Integrated Development Environment**
As identified by [Grossman 2007] the spreadsheet provides an integrated development environment which supports end users in their development of business critical applications. Support provided in terms of immediate feedback, error correction and debugging tools enables rapid prototyping and experimentation in the design and modification of spreadsheet models. It is key that development occurs within an already familiar environment for which users have a well-known mental model and which can be shared with other users.

**2.11 Ability to formalise a process**
Finally in the business context spreadsheets are often formalising a business process[Buckner 2004], for example submitting an expenses claim. Such models can be tailored specifically for the use case and importantly this can be done by the people who will be using the model who are empowered to evolve the model as needs change without the need for complex software projects.

**2.12 Conclusion**
The key strengths of the spreadsheet stem from its simple model of tables of cells containing data or formulas. This model has been grasped by a wide audience and used for a wide variety of purposes. Underlying this is the model's adaptability and flexibility, providing few constraints to usage in a common sharable format. Supporting this simplicity is the provision of a wide set of advanced features supporting more specialised tasks, these are provided on a continuum hiding advanced features from novice users presenting a simple interface. Finally the interoperability of these features combined with a wide user base has enabled the global success of the spreadsheet.

**3. CHALLENGES WITH CURRENT SPREADSHEET TECHNOLOGY**
While these features of spreadsheets have led its success, they have also caused a number of challenges. Some are known to daily spreadsheet users, while others are only widely known within professional and research communities. Addressing these challenges has been the subject of research for at least the past decade. Here we outline key challenges with spreadsheet technologies:

**3.1 Hidden Errors**
Humans are fallible creatures, hence it is widely known in the literature [Panko 1998][Powell 2008][Panko 2015] (though perhaps not in wider practice) that the error rate in the creation, evolution and use of spreadsheets is strikingly high. This is known to be the case not only in complex formulas but in data entry and use of many spreadsheet features. Overconfidence among users coupled with a lack of testing and the hidden nature of the structure of how formulas interact combine to result in a recurrent stream of horror stories in finance, industry[Thorne 2013] and research [Herndon 2014].

**3.2 Comprehensibility**
One underlying reason for this high error rate is known to be the user's difficulty in understanding the spreadsheet they are interacting with [Panko 2015] [Kohlhase et al 2015]. Formulas are hidden from view by default making understanding of logic, calculation and the flow of data through the model a challenge. Further it is a challenge to understand the wider hidden structure of the model, particularly when formulas interact across worksheets out of the immediate view of users. Such challenges mean that gaining a high level view of how a spreadsheet model works remains difficult [Hermans 2011].

**3.3 Complexity**
Spreadsheets have a tendency to grow over time, both in the amount of data stored and the complexity of the calculations performed (for an extreme example see [Birch et al 2014]). Without clear guidelines and a clear purpose the complexity of a spreadsheet is unbounded and challenging to deal with, particularly when such models are inherited by new users. There are few mechanisms to compartmentalise complexity within spreadsheets, such as the role functions, classes and modules provide within more general programming.

### 3.4 Adaptability and informality

One of the great strengths of spreadsheets is their adaptability however this causes a great weakness, spreadsheets often evolve to become unmanageable and error prone. With no rules or guidance for development of a model its complexity will tend increase and its integrity to degrade as the model is reused and adapted to new situations. This is particularly noticeable when a model is shared or inherited between multiple users. To address this a number of spreadsheet development guides[Grossman 2010] and processes [Ferreira and Visser 2012] have been developed though not widely adopted outside of their context.

### 3.5 Limited scale

Current spreadsheet technologies are unable to scale to support large datasets which are increasingly encountered. While the maximum number of rows Microsoft Excel supports has increased from 65536 ($2^{16}$) to 1,048,576 ($2^{20}$) over the last decade, this limit is still frequently less than the data encountered and which professionals seek to gain insight from.

### 3.6 Limited formal training

A large proportion of spreadsheet users are, beyond the basics, primarily self-taught, meaning that many will have picked up bad habits through lack of knowledge, or will be unaware of techniques that could help reduce their error rate or increase their efficiency.

### 3.7 Conclusion

Comprehending the complexity of a spreadsheet model to generate a good mental model of the structure of the spreadsheet remains the key challenge for professionals working with spreadsheets. Doing so while identifying and avoiding hidden errors in spreadsheets requires good training and the disciplined use of the adaptability and informality which is one the greatest strengths of the spreadsheet model. Augmenting this integrated development environment to assist management of this complexity remains a key challenge which will only grow as we approach scale limits of data supported by modern spreadsheets. This limitation is symptomatic of the one of the key Drivers of Change facing the spreadsheet – the rise of "Big Data".

### 4. DRIVERS OF CHANGE

The evolution of technology has and will continue to impact upon the evolution of the spreadsheet and affect their users. These drivers of change may impact gradually or bring rapid, fundamental change. It remains to be seen whether this will require augmentation of the spreadsheet as we know it with new features and tools or whether a more fundamental change of the underlying spreadsheet model will be required.

### 4.1 The rise of Big Data

It is now increasingly reasonable to collect and collate large corpuses of data which far exceed the ability of spreadsheets to store, manipulate or analyse. While common databases and specialist tools do exist for interacting with such data, these tools are beyond the experience or skill set of most spreadsheet users. This will decrease the effectiveness and the utility of spreadsheets in general practice, particularly as the opportunities and insight within large datasets are becoming increasingly valuable even for small companies. Analysing patterns in hundreds of thousands of historical orders or analysing millions of website traffic interactions are tasks increasingly common, even in small firms who will not have familiarity, expertise or access to more specialised tools than the humble spreadsheet.

### 4.2 The heterogeneity of data

As more data is collected, the variety of the data will correspondingly increase. There will be an increasing need to relate, fuse and analyse multiple datasets concurrently. This is frequently beyond the capability of spreadsheets and their lookup functions requiring tools traditionally only found in databases. Data will also no longer take traditional forms with new data including unstructured textual and image-based data. This data is beyond the scope of current spreadsheet technology.

### 4.3 Coping with unreliable data

Inevitably as the quantity of data to be processed grows additional challenges with its quality will be encountered. Current spreadsheets struggle to cope with malformed or inconsistent data and require many manual steps to reformat and coerce data into a clean state for working with. This assessment of the veracity of the underlying data is a key step for modern data science and will become increasingly important yet is hard to perform with existing spreadsheets.

**Table 1: The Future of Spreadsheets in the Big Data Era**

| The Success of the Spreadsheet | Drivers of Change | Directions of Change |
|---|---|---|
| Ubiquity | The rise of Big Data | Analysis of external Big Data |
| Unconstrained canvas | The heterogeneity of data | Heterogeneous unreliable data |
| An open box | Coping with unreliable data | Data fusion |
| Ability to communicate data | The rise of machine learning | Data analysis tools |
| Ability to store data in a computable format | The rise of Artificial Intelligence | Exploratory data visualisation |
| Lightweight databases | Rise of real-time data and the Internet of Things (IOT) | Cloud based analysis |
| Ability to model and test scenarios | | Improved Auditing tools |
| Accessible End-User Programming | The rise of cloud computing and Software as a Service | |
| Advanced Functionality | The rise of the App ecosystem | |
| Integrated Development Environment | Online collaboration suites | |
| Ability to formalise a process | Increased regulatory understanding of risk | **Research Directions** |
| | Enterprise spreadsheet management | Supporting End user Programming |
| **Challenges with Spreadsheets** | A more computer literate generation | Structuring of spreadsheets |
| Hidden Errors | A change of end user computing devices | Identification and auditing of errors |
| Comprehensibility | The rise of Business Intelligence software | Refactoring tools for spreadsheets |
| Complexity | The rise of data dashboard systems | Mechanisms for Code Reuse |
| Adaptability and informality | | Open source model repositories |
| Limited scale | | Dealing with Data |
| Limited formal training | | Data analysis techniques |
| | | Version control and collaboration in spreadsheets |
| | | Interacting with spreadsheets in new form factors |

### 4.4 The rise of machine learning

Supervised and unsupervised machine learning from data either in classification of data or the prediction of trends from historical data are becoming key techniques for the modern data scientist. Such methods are not found within modern spreadsheets, partly because they are unable to cope with the scale of data required for successful learning which will be an increasingly key technique.

### 4.5 The rise of Artificial Intelligence

The availability of larger volumes of data, computing power and machine learning techniques such as deep learning have empowered the rise of Artificial Intelligence which will impact upon the workplace in the coming years. Reports by [Frey & Osborne 2013] have suggested that many current administration and data processing jobs will be supported or replaced by AI systems in the future. Examples include the rise of digital assistances such as Amazon's Alexa, Microsoft's Cortana and Apples Siri. Such technology also appears in online shopping assistants or chat-bots. One example of the impact this may have is the use of an AI digital assistant to record the annual leave taken in an office, a role traditionally fulfilled by administration staff using a spreadsheet. This may greatly reduce the requirement for spreadsheet use. It may also be that digital assistants will help with the design of new spreadsheet models.

### 4.6 Rise of real-time data and the Internet of Things (IOT)

As surveyed by [Atzoria 2010] the Internet of Things is the convergence of several enabling hardware and communications technologies which enables the instrumentation of everyday situations and objects. The flow of information this provides typically exceeds the scale of spreadsheets to analyse. Such analysis is often time critical requiring the generation of real-time alerts, something beyond existing spreadsheets.

**4.7 The rise of cloud computing and Software as a Service**
Cloud computing has offered the opportunity to open traditionally closed business systems to the world online. Such systems are termed "Software as a Service" [Turner 2003] and present complex functionality (for example small business accounting) through simple websites with complex processing of data carried out within the cloud. This limits the need for the development and management of software and IT infrastructure. Many of the business roles which is now offered as "Software as a Service" would traditionally have been fulfilled by spreadsheets managed by professionals. While some systems enable using spreadsheets as the calculation engine of software as a service (e.g. Google Forms) this presents a challenge to the existing role of spreadsheets.

**4.8 The rise of the App ecosystem**
Many spreadsheets are created to fulfil specific purposes where no other software exists. The rise of the App eco-system on many platforms provides a wide range of niche applications. The challenge potentially facing spreadsheets is "Why create a spreadsheet model if "there's an app for that"".

**4.9 Online collaboration suites**
The rise of online office suites such as Google Docs and Office 365 is changing the nature of the way we use spreadsheets. These systems enable real time collaborative updates to a single document, they also enable the use of crowd sourcing of data and insight from a large number of participants. This is a new way of working with spreadsheets with multiple authors concurrently editing a document, this leads to more efficient collaboration and sharing of documents. However it also leads to challenges around ownership and the integrity of constantly evolving models which makes tracing the provenance of calculated information a challenge. Underling this is a version control, a widely known tool in the software development community but a newer challenge to spreadsheets.

**4.10 Increased regulatory understanding of risk**
Over the past 15 years a number of accounting scandals[Croll 2009] have been linked to errors in spreadsheet models. This combined with the Financial Crisis of 2008 has led to an increased focus on controlling risk within financial models, many of which are built as spreadsheets. This has led to a demand for auditing tools for spreadsheets in an attempt to alleviate the level of risk.

**4.11 Enterprise spreadsheet management**
Given this increased awareness of risk many enterprises are keen to take "control" over spreadsheet development and usage. This may take the form of an auditing team or specific training in spreadsheet development. Alternatively spreadsheets may be used to create the initial model or algorithm before being transformed by a development team into am authorised line of business application. Adapting spreadsheets to support this transition smoothly will be key to their utility in the future.

**4.12 A more computer literate generation**
Driven by the rise of code academies and the re-introduction of programming into the school curriculum in the UK the next generation of professionals will increasingly seek tools which enable them to fluently code bespoke solutions to their problems. This will increase use of advanced features of spreadsheets (for example macros and scripting) and simultaneously an abandonment of spreadsheets which are found to be "too simple" or "too constraining" for literate programmers.

**4.13 A change of end user computing devices**
The rise of mobile computing on tablets and smart phones and its new form-factor is now beginning to affect the workplace. Some office applications have been adapted to mobile phones with varying degrees of success as devices are generally optimised for consuming content rather than creating it. Spreadsheets are difficult to work with on such devices and this may affect their future usability.

**4.14 The rise of Business Intelligence software**
Such programs such as [Tableau, Spotfire, PowerBI] offer users a visual-first approach for exploring and analysing data by enabling large datasets to be graphed across many different data sources including database servers. Such systems are often used to create data dashboards.

**4.15 The rise of data dashboard systems**
One role traditionally performed by spreadsheets is the regular presentation of data as a dashboard of performance. This role, once the preserve of spreadsheets, is increasingly now filled by specialised systems, often implemented in a Software as a Service model which allow the creation of graphical dashboards updated in real time by querying database servers. Providing online or intranet dashboards these systems can be interrogated by a wider audience. While Microsoft's Excel has begun to move in this direction via the integration of dashboards in PowerBI the challenge to spreadsheets will increase.

**4.16 Conclusion**
These drivers present a threat to the dominance of the spreadsheet as a key tool for working with data and for modelling by future professionals. These threats take three forms:

1. The rise of increasingly available Big Data, its scale and variety combined with its unreliability mean that spreadsheets are no longer able to support it. Such data enables insightful methods of machine learning which are not supported by spreadsheets and which will power AI systems which may replace spreadsheets in some contexts.

2. A change in how spreadsheets are used, resulting in new demands. As seen in the rise of online collaboration and a change in the type and form factor of devices we use daily.

3. Specialised competitors to the spreadsheets including Business Intelligence suites for analysing and presenting data or on a wider scale the rise of App stores and Software as a Service systems offering fully featured software in cases where spreadsheets may once have been created. This is supported by an increased understanding of the risk of spreadsheet models and a desire for enterprise control over them.

These drivers of change present exciting opportunities to empower professionals with new tools and techniques to master the challenges of computing with data in the next decade. Alternatively new, more specialised, tools will be developed and mastered by (new?) professionals, this will lead to a side-lining of the venerable spreadsheet.

**5. DIRECTIONS OF CHANGE**
Given the drivers of change identified we now explore the likely directions of change in which spreadsheets are likely to evolve. It is unclear whether such evolution can naturally retain the underlying simplicity of the spreadsheet model or whether a revolution in the concept of a spreadsheet will be required. The authors believe that the in the future spreadsheets will support:

**5.1 Analysis of external Big Data**
As the volume of data increases spreadsheets will be forced to provide mechanisms for working with such Big Data. Given the scale of this data it will be necessary for it to be stored in external databases or files or APIs. Such data can then be queried, ingested, transformed and analysed through the spreadsheet engine in much the same way that large tables within spreadsheets are today. This trend is already in evidence with Microsoft's PowerBI tool suite [PowerBI]. This may also be powered by the idea of taking your algorithm to the data (residing in the cloud) rather than bringing the data to your calculation spreadsheet. This will require the ability to extract algorithms from spreadsheets.

**5.2 Heterogeneous unreliable data**
Spreadsheets will increasingly be called upon to analyse data types beyond structured numeric data. For example textual and unstructured document data, much of which is unreliable and has missing data points. As the volume of this data grows tools will be needed to deal with the transformation of such data with a better range of error states for the future spreadsheet.

### 5.3 Data fusion
While existing lookup functions provide the ability to relate datasets together within the spreadsheet there will be an increasing need to cross-reference and join external datasets together. This is one of the traditional strengths of SQL databases, though these tools are often too complex for most end users. Tools such as [Tableau], [Spotfire] and [PowerBI] are beginning to support this however native support within a spreadsheet would empower a wider audience.

### 5.4 Data analysis tools
Future spreadsheets must incorporate the new generation of data analysis tools, ranging from machine learning algorithms for classification of data or prediction to more advanced statistical tools for model testing. These tools need not only be applied in their traditional role but may be used to directly assist the spreadsheet user, one example of this trend is Excel's Flash Fill function which predicts additional or missing data items in a series.

### 5.5 Exploratory data visualisation
Traditionally data visualisation is used to communicate the results of analysis, a newer trend is to use interactive visualisation to explore large datasets identifying insight visually. This is evidenced by the success of tools such as [Tableau], [Spotfire] and [PowerBI]. Although these tools are excellent for exploring data they lack the flexibility for working and calculating with data which spreadsheets traditionally provide.

### 5.6 Cloud based analysis
The future of the spreadsheet will increasingly be affected by online collaboration, with concurrent editing of spreadsheets stored in a single authoritative repository. This presents the opportunity for the application of large amounts of computing to be applied to models for additional insight through techniques such as sensitivity analysis[Birch et al 2014]. Cloud based hosting will also enable models to be published to form the basis of an online application or service.

### 5.7 Improved Auditing tools
Given the increasing recognition of the danger and prevalence [Panko 2015] of spreadsheet errors combined with their hidden nature it is increasingly necessary for auditing tools to be developed and applied to spreadsheets. These may take the form of guidelines for modelling[Grossman 2010], or automated tools for detecting suspicious calculation formulas [Hermans 2012]. Further tools to support users with gaining an understanding of the structure of large models through techniques such as visualisation [Hermans 2011] should be employed. Underlying these all of these techniques are an adopting of standard techniques and tools for supporting programmers to the world of spreadsheets. It is very likely that this trend will continue as more techniques are usefully adopted into the end-user programming domain.

### 5.8 Conclusion
In light of the drivers of change identified the authors believe that the central direction of change in the future of the spreadsheet will be to aid spreadsheets in continuing to fulfil their daily role – dealing with data. As data grows in scale, becomes more heterogeneous and is less reliable better tools and support will need to be integrated into the spreadsheet model. The aggregation of cells into Table structures within a spreadsheet is the start of this trend, spreadsheets must continue to be expanded to work fluently with such Big Data, and perhaps other useful spreadsheet structures will emerge, for example "Table Summary", "Table Join", "Database Query ". Working with such data will drive a need to join or fuse datasets together, something which is traditionally tricky in spreadsheets via lookup functions and is the preserve of databases. Once spreadsheets are able to cope with such data tools will be needed to analyse and explore it. Machine learning for prediction of trends or classification of data is one technique well suited for bigger datasets which will need to be integrated into spreadsheets. Interactive visualisation of data for exploration is an increasingly common requirement and one that spreadsheets are beginning to adapt to[PowerBI]. The central challenge for the future will be how to adapt the spreadsheet model toward these directions without damaging the central successful paradigm of spreadsheets that make them so valuable to end users.

## 6. RESEARCH DIRECTIONS
Taking advantage of the opportunities presented for spreadsheets opens a number of fruitful research areas, among those identified we believe the most important are:

### 6.1 Supporting End user Programming
For most end users programming is a challenging prospect, spreadsheets provide a familiar context for understanding common concepts in programming and extending their knowledge. Supporting end users comprehension of spreadsheet models and programming concepts within is an active area of research[Hermans 2011]. Three key directions of research which assist in this are as follows:

### 6.2 Structuring of spreadsheets
Most spreadsheets have an internal structure containing tables, calculation areas and summaries, however few of these are treated as high level objects such as tables. Adding further high level structures to spreadsheets would improve the comprehensibility of models these might include external data imports, clusters of cells performing a calculation (the equivalent of a method in a programming language). One key missing component is the equivalent of a program class – a representation of an entity containing a group of properties about that object. Such entities may be discoverable within spreadsheet models[Abraham and Erwig 2006] and would help to structure models to the avoidance of errors.

### 6.3 Identification and auditing of errors
A current research direction which continues to bear fruit is the automatic identification of spreadsheet errors and poor practice[Aurigemma and Panko 2007] [Hermans 2012]. A number of research and commercial tools have been developed which support the auditing of spreadsheets. This research will continue to bear fruit in the future and will need to focus upon how data is transformed and processed in the age of Big Data. For example capturing and auditing the data pipeline within a spreadsheet - how data is imported and cleaned within a spreadsheet is likely to be a source of potential errors and useful tools.

### 6.4 Refactoring tools for spreadsheets
Having identified errors or quality issues it is common in programming development environments to provide tools to fix these issues. This maybe done either automatically or under the guidance of the user. While some research tools of this kind have been developed for spreadsheets[Abraham, Erwig 2007]they have not reached the end user. Further such tools should not only focused on the fixing of potential errors but on providing tools for transforming, or *refactoring*, the structure of spreadsheets. Potential refactoring's might include "standardise cell formatting", "split long formula" or "extract common calculation".

### 6.5 Mechanisms for Code Reuse
A major enabler of the development of complex programs are mechanisms for the reuse of complex functionality at various scales. Functions, Classes, Modules and Libraries of code all encapsulate specific functionality which may be reused again and again. Such mechanisms are not present within spreadsheets and their development would greatly enhance the scalability of complex spreadsheets and their rapid development by encapsulating key functionality. Challenges inherent within this are how to identify such commonality and how to extract it naturally. Extraction of such commonality will require refactoring tools and would enable repositories of shared functionality.

### 6.6 Open source model repositories
In recent years large open source repositories of code have appeared for most programming languages (for example Github). Such repositories may contain small functional modules, for inclusion in other systems, or complete complex programs. Surprisingly, given the size of the user group, there are few similar repositories for spreadsheets. It is possible that this is a cultural challenge, "my spreadsheet is confidential" or that the quality of an average model is poor or that spreadsheets are simply too bespoke for reuse. Alternatively it may be that spreadsheets lack the clear point of abstraction at which common functionality can be clearly extracted and made reusable by many people. Should this prove to be the case then the provision of such an abstraction should prove valuable by enabling the sharing of complex functionality which may be peer-reviewed to the avoidance of error and improvement of quality.

### 6.7 Dealing with Data

One of the biggest challenges facing spreadsheets is how to enable working with large amounts of data which exceeds that which may be usefully stored within a traditional model (is it useful to refer or view row 99,999,998?). Combining such data into the spreadsheet will be essential for its future as the scale of data we interact with daily grows considerable and is no longer capable of being sensibly stored locally. It may be that as with existing tools for processing Big Data, rather than bringing the data to the algorithm; the algorithm should be sent to the data. This enables distributed processing of large volumes of data on compute clusters with algorithms prototyped locally then sent to the data to be processed on a cluster. Further improvements will be necessary in the way spreadsheets deal with data, particularly when it is unstructured and unreliable; dealing with inconsistent data in spreadsheets at present requires a sharp eye, tools to highlight inconsistencies are being implemented but will need to improve to support a larger scale.

### 6.8 Data analysis techniques

Spreadsheets are a natural environment for processing and analysing data. However the tools available to process this data are limited. Future spreadsheets will need to be augmented with machine learning techniques and provide integration with stronger data processing tools and ecosystems such as Python and the statistical language R. Whether such complex tools can be integrated in a user friendly way remains to be seen but perhaps Visual Basic for Applications is not the language for the future of the spreadsheet? Perhaps a new domain specific language for spreadsheets should be developed?

### 6.9 Version control and collaboration in spreadsheets

Version control systems have proven essential in professional computer programming which enable large teams to work together working concurrently editing the same codebase. Each team member is able to work on different features, changes can be put through an automated and peer-review process as changes are merged either automatically or line by line in case of conflict. Recent versions of spreadsheets now incorporate a "Track Changes" functionality which enables highlighting of changes made by different users along with a comment and review system. Such tools are a start toward this but more robust version control systems are required particularly in the context of increasingly online and collaborative method of working where large teams interact with a single document concurrently.

### 6.10 Interacting with spreadsheets in new form factors

The form-factor of computers we use has barely changed in the office environment for several decades. We still expect spreadsheet users to have a mouse, a keyboard and a large high resolution screen. These assumptions are increasingly being challenged by the rise of mobile and tablet computing bringing new modes of interaction such as touch, stylus and voice control. None of these modalities are widely used in spreadsheets. Similarly the reduction in screen resolution makes using and creating large spreadsheets very difficult, particularly given the large grid like matrix. Exploiting these changes effectively within the spreadsheet requires further research into end-user studies and design. Further into the future the rise of Virtual and Augmented Reality will provide the spreadsheet with the potential to move into a natural 3d environment. Quite how the spreadsheet will look in this environment is an open research question!

### 6.11 Conclusion

Much of the recent research effort around spreadsheets focuses on the support of end users as they work with the existing spreadsheet model. Generally this is either through the identification of spreadsheet errors [Panko 2015] or with tools and techniques to support their understanding of the complexity of the spreadsheet. In the future research will be needed to adapt the spreadsheet to the challenges it faces as the environment in which spreadsheets are used changes. This will primarily be through an increase in the scale, variety and inconsistency of future datasets but also in the compute devices and online and more collaborative nature of future work. Evolving the spreadsheet to meet the challenges inherent in these changes provides an interesting research question, particularly if it is to be done without alienating existing spreadsheet users and the simple spreadsheet model which has been the underlying success of the spreadsheet over the past 30years.

## 7. IMPLICATIONS FOR USERS

Finally we explore the implications of the drivers of change for the end users of spreadsheets. Future users should be increasingly technically literate having grown up in an online world using smartphones and tablets from a young age and being taught computer science in school[Brown et al 2014]. The authors believe that the following are some of the changes users should expect:

### 7.1 The same model

Fundamentally users should expect little change to the basic spreadsheet model which has stood the test of time and proved its versatility. Indeed changing this fundamental models is liable to encounter substantial opposition and the inertia effect of requiring retraining of a large number of users. This should not be surprising as the fundamental reasons for which spreadsheets are used daily are not going to change – the recording and manipulation of data, the creation of a calculation model.

### 7.2 Better Support

A growing body of research is focused on the support of end users of spreadsheets who should expect to see this work support them in their daily use of spreadsheets. From the identification of errors to the extraction of structure within spreadsheets these tools will increasingly aid users in working efficiently and more importantly safely with spreadsheet models. This will take the form of tools for gaining a better understanding of the spreadsheet model, perhaps through visualisation, and by tools for auditing existing models, suggesting fixes to identified issues. This trend should result in an improvement in the quality of spreadsheet models.

### 7.3 Easier ways of working with data

It is clear that the scale and variety of data which future users will seek to work with will continue to grow. Future spreadsheets must support users in this or they will lose their relevance to the next generation of users. This may be done by enabling spreadsheets to import large quantities of external data or by enabling the algorithms within spreadsheets to be executed on the cloud. At the least spreadsheets will require a more consistent data processing pipeline recording the steps taken to clean a data table or enabling better fusion or joining of datasets. As the number of datasets grows tools for interactively exploring such data will be provided as evidenced by Microsoft's PowerPivot [Ferrari and Russo 2013]. Users should further expect to have a wider range of tools for analysing such data, machine learning is proving a key technique in analysing Big Data and will likely need to be incorporated into the modern spreadsheets expanding suit of tools.

### 7.4 Changing ways of working

Over the past 20 years we have seen the rise of laptop's, smart phones, tablets and online collaboration suites. Each of these has changed the way we work and is having an effect on the way we interact with the humble spreadsheet. Future trends will likely see the rise of virtual and augmented reality which may impact upon the spreadsheet (which may no longer be limited to 2 dimensions!). In addition we will see a more collaborative online approach to working with spreadsheets with multiple authors and a more structured approach to model sharing and publishing. Perhaps we will see a rise of "open source" spreadsheet repositories or components.

### 7.5 Conclusion

It remains to be seen whether these changes will be incorporated directly into spreadsheet software and the essential spreadsheet model or whether they will be evolved in external complimentary tools.

It is likely that the future spreadsheet user will be offered a more extensive toolset to augment their basic spreadsheet, the challenge for such a user will be to navigate and learn these tools and to use them effectively. Conditions for using spreadsheets will also vary from different computing form factors to a more online and collaborative setting. This will present new opportunities and challenges for users who should still expect to find spreadsheets a key tool for working with data and modelling.

## 8. CONCLUSIONS

The success of the spreadsheet stems from its simple underlying model and the adaptability of that model to the everyday tasks which millions of people carry out. While this model and its implementations do suffer from some challenges (notably the hiddenness of errors) its continued success argues well for the future of the spreadsheet, particularly as this is driven by a continued underlying business need.

> *"...spreadsheets will **always** fill the void between what a business needs today and the formal installed systems..."*
> *[Glass 2009]*

The future of the spreadsheet will be driven by how they meet the Drivers of Change identified in this paper. These form challenges in two directions, changes in the nature of the tasks spreadsheets are used for and changes in the context in which they are used. Firstly the tasks spreadsheets are used for will change as the data we encounter daily grows in scale, heterogeneity and unreliability. We will also seek to use spreadsheets for different purposes in light of such data, either to explore it statistically or visually or to fuse or process it into new forms. Similarly new tools will be required such as machine learning as it becomes an increasingly effective method of gaining insight as the scale of data grows. Secondly the context in which spreadsheets are used will change either in the type of computing device used or toward a more social collaborative online spreadsheet experience. With an increased understanding of the risk of spreadsheets enterprise control (especially in a cloud based environment) will change the way we interact with spreadsheets with increased auditing and tool support. This does present opportunities for spreadsheets, for example the publishing of models to open source repositories or as the basis of a "Software as a Service" model.

Recent developments in spreadsheets in the past decade have primarily been augmentation of the successful core spreadsheet concept with advanced tools and functionality to support new needs of users. It remains to be seen whether simply augmenting the tools provided to users without adapting the core spreadsheet concept will be sufficient to ensure the success of the spreadsheet in the coming decades. This is an exciting and fruitful area of research as we seek to better understand and support millions of end users in their computing with spreadsheets.

## ACKNOWLEDGEMENTS

The authors would like to thank ICAEW and Imperial College London's Data Science Institute for hosting the Future of the Spreadsheet workshop and all of the conference participants for their insight and lively discussion, it is only with a strong expert community that the spreadsheet has a future.